\documentclass{emulateapj}
\usepackage{graphicx}
\usepackage[usenames,dvips]{color}
\usepackage{color}
\usepackage{soul}

\def\arcsec{$\,^{\prime\prime}$~}

\newcommand{\be}{\begin{equation}}
\newcommand{\bel}[1]{\begin{equation}\label{eq:#1}}
\newcommand{\ee}{\end{equation}}
\newcommand{\bd}{\begin{displaymath}} 
\newcommand{\ed}{\end{displaymath}}   
\newcommand{\bea}{\begin{eqnarray}}
\newcommand{\beal}[1]{\begin{eqnarray}\label{eq:#1}}
\newcommand{\eea}{\end{eqnarray}}

\newcommand{\eqref}[1]{\ref{eq:#1}}

\newcommand{\lsim }{{\lower0.8ex\hbox{$\buildrel <\over\sim$}}}
\newcommand{\gsim }{{\lower0.8ex\hbox{$\buildrel >\over\sim$}}}

\def\aap{ A\&A}

\def\apjs{ApJ Supp}

\def\Chandra{${\it Chandra}$}

\def\simge{\mathrel{%
   \rlap{\raise 0.511ex \hbox{$>$}}{\lower 0.511ex \hbox{$\sim$}}}}
\def\simle{\mathrel{
   \rlap{\raise 0.511ex \hbox{$<$}}{\lower 0.511ex \hbox{$\sim$}}}}

\newcommand{\Msun}{\ifmmode {M_{\odot}}\else${M_{\odot}}$\fi}
\newcommand{\Lsun}{\ifmmode {L_{\odot}}\else${L_{\odot}}$\fi}
\newcommand{\Rsun}{\ifmmode {R_{\odot}}\else${R_{\odot}}$\fi}

\shorttitle{Thermal X-ray Radiation from SAX J1808.4-3658}
\shortauthors{Heinke et al.}

\begin{document}
\title{Further Constraints on Thermal Quiescent X-ray Emission from SAX J1808.4-3658  $^1$}  

\author{C.~O. Heinke\altaffilmark{2,3,4}, P.~G. Jonker\altaffilmark{5,6}, R. Wijnands\altaffilmark{7}, C.~J. Deloye\altaffilmark{4}, and R.~E. Taam\altaffilmark{4,8}}

\altaffiltext{1}{Based on observations obtained with XMM-Newton, an ESA science mission with instruments and contributions directly funded by ESA Member States and NASA.}
\altaffiltext{2}{University of Alberta, Dept.\ of Physics, 11322--89 Avenue, Edmonton, AB, T6G 2G7, Canada; cheinke@phys.ualberta.ca}
\altaffiltext{3}{University of Virginia, Dept.\ of Astronomy, PO Box 400325, Charlottesville, VA 22902}
\altaffiltext{4}{Northwestern University, Dept.\ of Physics \&
  Astronomy, 2145 Sheridan Rd., Evanston, IL 60208}
\altaffiltext{5}{SRON, Netherlands Institute for Space Research, Sorbonnelaan 2, 3584~CA, Utrecht, the Netherlands; }
\altaffiltext{6}{Harvard--Smithsonian  Center for Astrophysics, 60 Garden Street, Cambridge, MA~02138, MA, U.S.A.}
\altaffiltext{7}{Astronomical Institute "Anton Pannekoek", University of Amsterdam, Kruislaan 403, 1098 SJ, The Netherlands}
\altaffiltext{8}{ASIAA/National Tsing Hua University - TIARA, Hsinchu, Taiwan}


\begin{abstract}

We observed SAX J1808.4-3658 (1808), the first accreting millisecond pulsar, in deep quiescence 
with XMM-Newton and (near-simultaneously) Gemini-South.  
The X-ray spectrum of 1808 is similar to that observed in quiescence 
in 2001 and 2006, describable by 
an absorbed power-law with photon index $1.74\pm0.11$ and 
unabsorbed X-ray luminosity $L_X=7.9\pm0.7\times10^{31}$ ergs s$^{-1}$, 
for $N_H=1.3\times10^{21}$ cm$^{-2}$.  
Fitting all the quiescent XMM-Newton X-ray spectra with a power-law, 
we constrain any thermally emitting neutron star with a 
hydrogen atmosphere to have a temperature less than 30 eV and 
$L_{NS}$(0.01-10 keV)$<6.2\times10^{30}$ ergs s$^{-1}$.  
A thermal plasma model also gives an acceptable fit to the continuum.  Adding a neutron star component to the plasma model produces less stringent constraints on the neutron star; a temperature of 36$^{+4}_{-8}$ eV and $L_{NS}$(0.01-10 keV)$=1.3^{+0.6}_{-0.8}\times10^{31}$ ergs/s.
In the framework of the current theory of neutron star heating and cooling, the constraints on the thermal luminosity of 1808 and 1H 1905+000 require strongly enhanced cooling in the cores of these neutron stars.  

We compile data from the literature on the mass transfer rates and quiescent thermal 
flux of the largest possible sample of transient neutron star LMXBs.  
We identify a thermal component in the quiescent spectrum of the accreting millisecond 
pulsar IGR J00291+5934, which is consistent with the standard cooling model.  
The contrast between the cooling rates of IGR J00291+5934 and 1808 
suggests that 1808 may have a significantly larger mass. 
This can be interpreted as arising from differences in the binary evolution 
history or initial neutron star mass in these otherwise similar systems.  
 
\end{abstract}


\maketitle

\section{Introduction}\label{s:intro}

The X-ray transient SAX J1808.4-3658 (hereafter 1808) has provided
many fundamental breakthroughs in the study of accreting neutron
stars (NSs). It was discovered in 1996 by {\it BeppoSAX}'s Wide Field
Cameras, and type I X-ray bursts were seen, identifying it as an
accreting NS and constraining the distance 
\citep{intZand98,Galloway06}.  Coherent millisecond X-ray
pulsations, the first discovered in accreting systems,
 were identified during an outburst using the Rossi X-ray Timing 
Explorer \citep[RXTE;][]{Wijnands98}.  
Burst oscillations have also been seen at 1808's 
401 Hz spin frequency, confirming that thermonuclear burst
oscillations in low-mass X-ray binaries (LMXBs) represent the spin
period of the NS \citep{intZand01b,Chakrabarty03}.
 
 Transiently accreting NSs in quiescence are seen to have a soft,
blackbody-like X-ray spectral component, and/or 
a harder X-ray component generally fit by a power-law of photon index
1-2 \citep{Campana98a}.  The harder component is of unknown origin;  
an effect of continued accretion or a shock from a pulsar wind have 
been suggested \citep{Campana98a}.   
 The blackbody-like component is generally understood as the 
radiation of heat from the NS surface. This heat is produced by nuclear fusion in the deep crust during accretion, and is radiated from the surface  
on a timescale of $10^4$ years, producing a steady
quiescent thermal NS luminosity \citep{Brown98, Campana98a, Haensel90}. 
The deep crustal heating rate can be computed if the mass transfer rate is 
known (or estimated). 
Some transiently accreting NSs have been shown to have very low 
quiescent thermal X-ray luminosities, indicating either 
enhanced neutrino emission or mass transfer rates much lower than have been previously inferred \citep[e.g.][]{Wijnands01,Jonker04,
Tomsick04,Jonker06}.  
The coolest of these provide the strongest constraints 
to date on neutrino cooling from NS cores, as a broader range of cooling 
rates is required from X-ray transients than from young cooling pulsars 
\citep[cf. ][]{Page04,Yakovlev04}. 

X-ray observations in quiescence have shown 1808 to have one of the lowest quiescent thermal luminosities yet measured from any 
accreting neutron star \citep{Campana02,Heinke07a,Jonker07a}. 
1808 is of particular importance because of its well-known distance \citep{Galloway06} and relatively stable (within a factor of 2) time-averaged mass transfer rate onto the NS measured over multiple outbursts.  
1808's low quiescent thermal luminosity indicates that most of the heat absorbed by the NS core during accretion is reradiated not as thermal X-ray emission, but through neutrino cooling processes \citep{Yakovlev04}.

We have obtained a third XMM observation of 1808 
in quiescence in 2007, in conjunction with near-simultaneous 
(separated by 6.5 hours) Gemini optical ($g'$ and $i'$) imaging.  
The science goals included further constraining the thermal component of the X-ray emission, constraining X-ray variability in quiescence, and measuring the sinusoidal orbital optical modulation nearly simultaneously with an X-ray observation to determine the origin of the optical modulation.  
One key finding is that SAX J1808.4-3658 provides one of the two most 
constraining upper limits on the thermal component for any accreting neutron star. Accordingly, 
we report on the X-ray analysis here, along with comparison to other X-ray 
transients in quiescence, while the optical analysis and comparison of the X-ray and 
optical results are presented in a companion paper \citep{Deloye08}.

\section{Data Reduction}\label{s:obs}

We observed 1808 on March 10-11, 2007 (ObsID 0400230501; starting at 16:24 UT) 
for 49 ksec with XMM's EPIC camera, using two MOS CCD detectors 
\citep{Turner01} with medium filters and one pn CCD detector 
\citep{Struder01} with a thin filter.   
All data were reduced using FTOOLS and SAS version {7.0.0}.   
Soft proton flares were excluded by excluding times when the 
total MOS count rate exceeded 4 0.2-12 keV counts per second, and times when 
the total pn count rate exceeded 20 0.2-12 keV counts per second. 
This left 36.9, 49.7, and 49.8 ksec in the 2007 pn, MOS1, and MOS2 datasets.
Event grades higher than 12 were also excluded.  We
extracted spectra from a 10\arcsec circle around the position of 1808, 
 and combined the pair of simultaneous MOS 
spectra and responses using FTOOLS.  We generated 
response and effective area files using the SAS tasks {\it rmfgen} and
{\it arfgen}, and produced background spectra from  90\arcsec
circular source-free regions on the same CCD. 
The spectra were grouped to $>$15 counts per bin for the MOS data, 
and $>$30 counts per bin for the pn data (other choices gave similar results). 

\subsection{X-ray Variability}

We produced  
background-subtracted lightcurves of the 2007 pn data within 
SAS, and analyzed them using HEASARC's  
XRONOS software \footnote{http://heasarc.gsfc.nasa.gov/docs/xanadu/xronos/xronos.html}.
Kolmogorov-Smirnov and $\chi^2$ tests on the first 37 ksec of 0.2-12 keV pn data 
(mostly unaffected by background flaring) revealed mild evidence of variability,  
as the probability of a constant flux is $3\times10^{-2}$ and $3\times10^{-3}$ for the two tests respectively.

\section{X-ray Spectral Analysis}\label{s:spec}

Our X-ray spectral analysis includes photoelectric absorption (XSPEC model 
{\it phabs}), with a hydrogen column density, $N_H$, fixed 
at the interstellar value of $1.3\times10^{21}$ 
cm$^{-2}$ \citep{Dickey90}.
We checked this $N_H$ value by analyzing a series of Swift observations taken during the tail of 1808's 2005 outburst \citep{Kong05,Campana05a}.  
We used window timing data from June 17, 23, and 29, and July 7 and 13, and photon counting data from June 17 and 23.  
The June 17 observation suffered from pileup in photon counting mode.  We addressed this by excluding the central 18'' of the point-spread-function (chosen by fitting a King model to the radial profile \footnote{Following http://www.swift.ac.uk/pileup.html.}).  
An absorbed power-law model (with only $N_H$ fixed between observations) gave a nearly reasonable fit to the 7 Swift spectra ($\chi^2_{\nu}$=1.19, null hypothesis probability=1.4$\times10^{-4}$), while an absorbed power-law plus blackbody model gave a slightly better fit ($\chi^2_{\nu}$=1.15, null hypothesis probability=1.8$\times10^{-3}$).  Photon indices ranged from 1.8 to 2.3 for the simple power-law fits, or 2.2 to 2.4 for the power-law plus blackbody fits, while the blackbody temperatures ranged from 0.7 to 1.2 keV.  The best fit $N_H$ for either fit was $1.2\pm0.1\times10^{21}$ cm$^{-2}$, consistent with the Dickey value.\footnote{After this paper was submitted, an independent analysis of the Swift data found $N_H=1.3\pm0.1\times10^{21}$ cm$^{-2}$ \citep{Campana08a}.}
We also tested models with photoelectric 
absorption as a free parameter, finding $N_H$ consistent with 
the outburst value.  Quoted errors are at 90\% confidence. 

We simultaneously fit our pn and MOS spectra along with spectra from the 
two previous XMM observations from March 24, 2001 and Sept.~14, 2006 
\citep[see ][for details of these observations]{Heinke07a}.  
The X-ray characteristics are similar to those observed in 
the prior observations \citep{Campana02}; the spectrum can be well fit 
with a power-law of photon index $1.74\pm0.11$, while a 
hydrogen-atmosphere model (the 
NSATMOS\footnote{http://xspec.gsfc.nasa.gov/docs/xanadu/xspec/models/nsatmos.html}  
model of \citet{Heinke06a}, or the 
similar NSA model of \citet{Zavlin96}) gives very poor fits.  

Although each of the spectra are well-fit by an absorbed power-law with 
the $N_H$ observed in outburst, requiring all spectral parameters to be 
identical across observations produces somewhat poor fits 
($\chi^2_{\nu}=1.09$, null hypothesis probability or nhp=0.03).  
Allowing a constant normalization to vary between the observations, 
we find that the 2001 flux is 1.29$^{+.23}_{-.21}$ of the 2006 flux,
 and the 2007 flux is 1.46$^{+.20}_{-.17}$ of the 2006 flux.  
This suggests that at its lowest flux levels, 1808 remains 
X-ray variable in quiescence \citep[see also][]{Campana08a}.  Such variability could indicate that the 
X-ray emitting process is powered by time-variable accretion, either 
onto the NS or in the interaction of a disk and the NS magnetosphere.

\begin{figure}
\figurenum{1}
\includegraphics[angle=270,scale=.35]{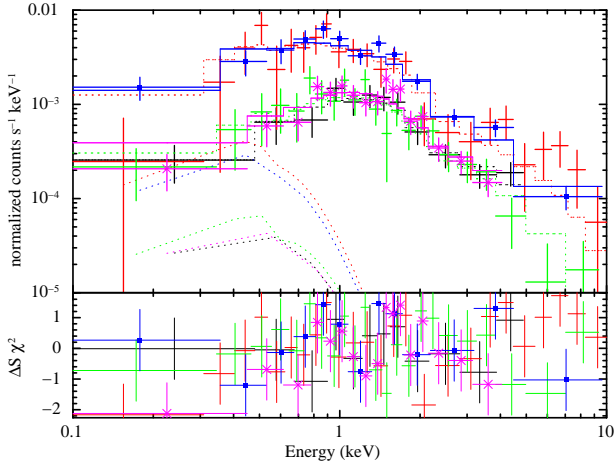}
\caption[1808_nspow.ps]{ \label{fig:spectrum}
 XMM X-ray spectra (data and best-fit power-law model) of 1808.  
Solid (blue) line with filled squares: 2007 pn data, model.  
Solid (magenta) line with crosses: 2007 MOS data, model. 
Other data is from the 2001 MOS (black), 2006 MOS (green), and 2006 pn (red) XMM observations 
\citep{Heinke07a}. 
The maximum acceptable contribution (90\% conf.) from the NSATMOS component is indicated by a dotted line, prominent at low energies.
(See the online edition of the Journal for a color version of this figure.)
} 
\end{figure}

Allowing only the $N_H$ to vary improves the fit only slightly (nhp=0.04), and 
we note that a different $N_H$ in outburst than quiescence is rarely seen in quiescent 
LMXBs, especially those like 1808 at relatively low inclination \citep{Jonker04c}.  
We fix the $N_H$ at $1.3\times10^{21}$ cm$^{-2}$ and free the power-law photon 
index and normalizations, including also a NSATMOS component with 
NS mass=1.4 \Msun and radius 10 km, with the same temperature between observations. 
We freeze the distance to 1808 at 3.5 kpc, as measured by \citet{Galloway06}. 
 The parameters of this fit are listed in Table 1, and it is shown in Figure \ref{fig:spectrum}.  
No thermal component is required, but a thermal component with $kT<$30 eV 
(90\% confidence) is permitted, thus placing a limit on the NS's 
thermal bolometric 
(0.01-10 keV) luminosity $L_{NS,bol}<6.2\times10^{30}$ erg s$^{-1}$.  
This is a substantially tighter constraint on the NS's quiescent thermal luminosity than that of \citet{Heinke07a}, and may be the tightest constraint on the thermal bolometric luminosity of a NS in an X-ray transient.  \citep[Note that 1H1905+000 has a much lower {\it total} X-ray luminosity than 1808; see][]{Jonker07a}.  This tighter constraint is produced by doubling the XMM pn exposure, and thus nearly doubling our sensitivity.  

The rather tight distance limits of \citet{Galloway06} ($3.5\pm0.1$ kpc) 
produce only a 6\% uncertainty in our upper limit.  Increasing the assumed $N_H$ by $10^{20}$ cm$^{-2}$ increases this limit by $\sim20$\%; decreasing the $N_H$ decreases the limit similarly.  Changing the assumed NS mass to 2 \Msun (as suggested by the results of Deloye et al. 2008) increases the upper limit by 10\%, as does altering the assumed NS radius to 13 km.

The nature of the high-energy spectral component in qLMXBs is unclear, and thus other continuum models are possible.  
One possibility for the emission is a hot plasma, possibly a shock 
between an infalling accretion stream and a pulsar wind 
\citep[e.g.][]{Campana02,Bogdanov05}.  
Such a model is not ruled out by the spectral fits, which allow 
($\chi^2_{\nu}=0.95$ for 84 degrees of freedom) a hot 
plasma MEKAL model with $kT=$4-8 keV instead of a power-law spectrum. 
Adding a faint NSATMOS component is modestly preferred in this spectral fit; 
the best fit has $kT=36^{+4}_{-8}$ eV, for a bolometric NS luminosity of $L_X=1.3^{+0.6}_{-0.8}\times10^{31}$ ergs/s (Table \ref{tab:1808spec}).  
This model allows a larger NS component due to the flatter shape of the model at low energies, and demonstrates the dependence of our constraints on the continuum model.  
The nature of the hard spectral component, and thus choice of MEKAL vs. power-law model, could be tested by a sensitive search for line emission, or by measuring the hard tail out to higher X-ray energies, with next-generation X-ray instruments.  

\section{Comparison to other results}

\citet{Heinke07a} noted a possible trend of NSs with low mass transfer rates having faster cooling.  
To test this, we have increased our sample of LMXBs with useful constraints on their mass transfer rates and quiescent thermal luminosities (Table \ref{tab:mdots}).  
For several systems, we take both values from the recent literature.  
For some, we compute the quiescent 0.01-10 keV thermal NS luminosity given the 0.5-10 keV fluxes in the literature and NS atmosphere models, or by reanalyzing \Chandra\ or XMM data on quiescent NSs.  
For several systems, we estimate the time-averaged mass transfer rate (or an upper limit) from X-ray flux histories in the literature, since the RXTE ASM lightcurves \citep[used in ][]{Heinke07a} generally do not cover their full outbursts.  Distance estimates are often rather poorly quantified.  We take best estimates of these distances from the quoted references, noting that an uncertainty of 50\% in distance is a reasonable upper bound.  Changes in distance at this level will not greatly affect our (broad) conclusions.  We also list orbital periods where known.  
We discuss a few systems in detail below.  

\subsection{Comparison to other LMXBs in quiescence}

\citet{Jonker07a} obtained a long observation of the quiescent NS LMXB 1H1905+000, deriving a very tight limit on the quiescent bolometric NS flux of $<1\times10^{31}$ ergs/s (for a 0.1 keV blackbody).   The mass transfer rate of 1H1905+000 is not known; but for any mass transfer rate $>10^{-12}$ \Msun/yr, the quiescent luminosity limit is below the fiducial pion cooling curve of \citet{Yakovlev04}.  On the other hand, the mass transfer rate for 1808 is rather well-known, as the mass transfer rate can be inferred from RXTE ASM monitoring of multiple outbursts, and by predictions from gravitational radiation \citep{Bildsten01}, which agree very nicely.  

\citet{Jonker07b} used \Chandra\ to identify the likely quiescent X-ray counterpart to 1M1716-315, with a soft thermal spectrum.  
We assume H-poor material, as suggested by \cite{Jonker07b}, to select the larger distance of \cite{Jonker04}).  
We utilize the historical X-ray fluxes of \citet{Jonker07b} to estimate an average mass transfer rate of \.{M}$\sim2.5\times10^{-10}$ \Msun/year over 37 years, which we take as an upper limit.

\citet{Tomsick07} identified the quiescent X-ray counterpart to 4U1730-22. We use the outburst history in \citet{Chen97} to compute its time-averaged mass transfer rate over the past 37 years.  

4U 2129+47 was a bright X-ray source from 1971 until sometime between 1980 and 1983, and has been in quiescence since then \citep{Pietsch86}.  4U 2129+47 is optically bright enough ($B\sim16.8-18.3$) that historic photographic plates are useful; they show 4U 2129+47 to be ``on'' from 1963 to 1979, and ``off'' from 1938 to 1943 \citep{Wenzel83}.   We estimate an upper limit for the average X-ray flux of $F_X\sim8\times10^{-10}$ ergs/s (0.1-20 keV), maintained over 20 of the last 45 years.  Using a distance of 6.3 kpc \citep{Cowley90}, we derive an upper limit of  $L_X<4\times10^{36}$ ergs/s.  However, the lightcurve in outburst shows extended V-shaped eclipses, indicating that we observe only scattered light from an accretion disk corona.  This may decrease our observed flux, possibly by a factor of more than 10.\footnote{\citet{Jonker01} estimate that the luminosity of the accretion disk corona source 2A 1822-371 is reduced by a factor of 40 due to its high inclination.}
 The peak luminosity of a non-radius-expansion X-ray burst observed by \citet{Garcia87} is $L=5\times10^{36}$ ergs/s.  This is a factor of 40 lower than the Eddington limit for hydrogen-rich material.  Allowing for a factor of 40 reduction in our observed flux, we estimate a (conservative) upper limit of $L_X<1.6\times10^{38}$ ergs/s, and \.{M}$<5.2\times10^{-9}$ \Msun/year.  

\subsection{Comparison to other accreting MSPs in quiescence}

Of the other accreting millisecond X-ray pulsars, deep quiescent studies have not been performed for Swift J1756.9-2508 and HETE J1900.1-2455.  For most of the remainder, we compute mass transfer rates using the distance estimates, time-averaged bolometric fluences, and recurrence time limits in \citet{Galloway06b}.  For XTE J1814-338, we assume a recurrence time of 19 years \citep{WijRey03}.  
For XTE J1751-305, the discovery outburst seems to have been much brighter than the other three recorded outbursts \citep{Markwardt02,Linares07}, so we add the estimated outburst fluences and average over the past 12 years. 

XTE J1814-338 (1814) does not have a published quiescent flux measurement, but one archival 
(Sept.\ 6, 2005) XMM observation exists.  We have analyzed this dataset to search for 1814 
in quiescence.  We excise periods of high background, $>$30 ($>6$) cts/s in the 0.2-10 keV pn 
(MOS) data, giving 22.9, 30.9, 29.5 ksec of pn, MOS1, and MOS2 data respectively.  A slight 
flux enhancement may be seen in the 0.2-4 keV images, although it is not highly significant.  
We estimate the flux of 1814 by computing the counts within 15'' of 1814's nominal 
position, and subtracting background estimated from an annulus from 20'' to 40'' away.  
Calculating the fluxes for an absorbed power-law of photon index 2 using PIMMS gives an 
(averaged) $F_X$(0.5-10) of $6.3\pm4.3\times10^{-15}$ ergs cm$^{-2}$ s$^{-1}$.  Using a 
blackbody of temperature 0.2 keV gives $F_X=3.9\pm1.6\times10^{-15}$ ergs cm$^{2}$ s$^{-1}$.  
We therefore identify $F_X<9\times10^{-15}$ ergs cm$^{-2}$ s$^{-1}$ as a 3$\sigma$ upper limit 
on the blackbody-like flux\footnote{Since a harder spectrum gives a higher upper limit, the total luminosity could be slightly higher.} from a NS in 1814, and for an assumed distance of 8 kpc, $L_{NS}$(0.01-10 keV)$<1.7\times10^{32}$ ergs/s. 

IGR J00291+5734's (hereafter 00291) quiescent X-ray flux has been observed with \Chandra\ on five occasions, reported by \citet{Jonker05}, \citet{Torres08}, and \citet{Jonker08}.  The third observation found 00291 to be significantly brighter (factor of two), with a softer spectrum, than the first two or the later observations \citep{Jonker05}.  
This observation can be fit with a NS atmosphere model alone, but the implied radius is rather small (only $3^{+2}_{-1}$ km for a distance estimate of 4 kpc; see Galloway et al. 2005, Torres et al. 2008, Jonker et al. 2005).  However, the addition of a power-law component, dominant at high energies, allows a reasonable NS radius.  The temperature of the NS appears higher in this observation than in other observations.  Exploring the nature of this variation lies outside the scope of this work.  Since only the faintest blackbody-like flux measurement can represent the emission from a slowly cooling NS heated by multiple outbursts, this observation is in any case not relevant to our purpose.

\begin{figure}
\figurenum{2}
\epsscale{0.6}
\includegraphics[angle=270,scale=.35]{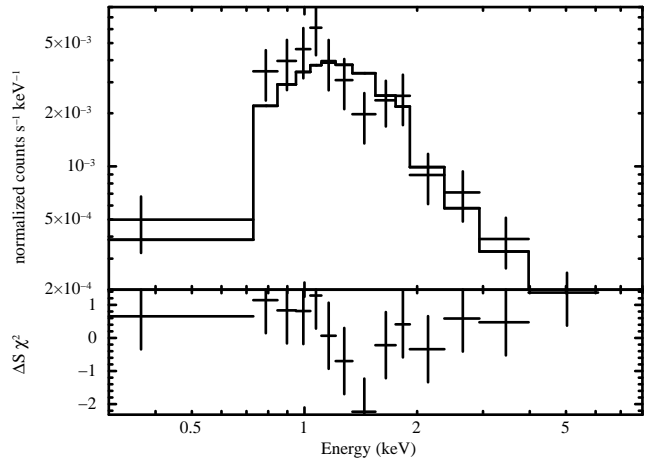}
\caption[IGR_justpow.ps]{X-ray spectrum of IGR J00291+5934, fitted with a power-law alone.  Data and model in upper panel, residuals in lower panel.
}\label{fig:00291pow}
\end{figure}

\begin{figure}
\figurenum{3}
\epsscale{0.6}
\includegraphics[angle=270,scale=.35]{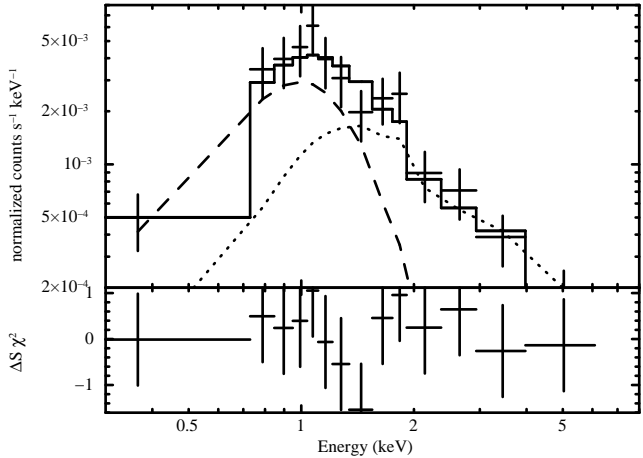}
\caption[IGR_nspow_bin10.ps]{X-ray spectrum of IGR J00291+5934, fitted with a power-law and NS atmosphere model.  The NS atmosphere component is shown by the dashed line, and the power-law component by the dotted line, in the upper panel, while residuals are shown in the lower panel.  Note the improvement in the fit compared to Figure \ref{fig:00291pow}.
}\label{fig:00291nspow}
\end{figure}

The other four observations of 00291 show a similar spectrum and X-ray flux (of $7\times10^{-14}$ ergs cm$^{-2}$ s$^{-1}$, 0.5-10 keV, for a power-law fit).  We analyze the longest observation, from 2005 November (it has 143 counts, vs. 36, 63, and 22 from the others), with slightly different methods than those used by \citet{Torres08}.  
We fit a model consisting of an absorbed powerlaw, with an optional NS atmosphere, with $N_H$ fixed to $4.6\times10^{21}$ \citep[the average value from several methods summarized by ][]{Torres08}.  We binned the source spectrum so that each bin contains at least 10 counts (using 15 or 20 counts/bin yields 
similar results).  We set the NS distance to 4 kpc \citep{Galloway05}, radius to 10 km, and  
mass to 1.4 \Msun.  We find that the power-law component is absolutely required, but that the NS component significantly improves the fit (see Table \ref{tab:00291spec}, Figures \ref{fig:00291pow} and \ref{fig:00291nspow}).  
The fraction of the 0.5-10 keV flux attributable to the NS component is 44$^{+22}_{-27}$\%. 

The reduced $\chi^2$ is reduced by half with the addition of the NS component, and the temperature of the NS component is inconsistent with the model minimum (effectively zero) at the $>$99\% level ($\Delta\chi^2$=7.0).
An F-test indicates that the probability of attaining an equivalent fit improvement by adding one model parameter is only 0.4\%. 
\citet{Protassov02} showed that the F-test is often inaccurate for testing the necessity of adding an additional spectral component.  We therefore simulated 100 datasets using the best-fit absorbed power-law model, and fit them with this model and with the absorbed NS plus power-law model.  None of our simulations gave a $\Delta \chi^2$ or F-statistic larger than that produced by our model, allowing us to conclude that the probability of concluding incorrectly that a NS component is required for 00291 is less than 1\%.

The difference between our results and those of \citet{Torres08} (which found an upper limit of 19\% to a thermal component for 00291) may be attributed to four factors: our inclusion of the lowest energy bin in our fit (this bin is dominated by data above 0.5 keV), our use of a fixed distance (which imposes a relation between NS temperature and NS flux), Torres et al's use of a fixed NS temperature, and, perhaps most importantly, Torres et al.'s method of first fitting an absorbed power-law, and then fixing the best-fit power-law parameters before adding and constraining a blackbody component.  This method is likely to underestimate the flux that may be present in a second component, since the parameters of the first component are not allowed to vary from those of the best one-component fit.  We have replicated the spectral fit of Torres et al., with similar results (the NS is $<19$\% of the 0.5-10 keV unabsorbed flux), but when we free the power law component parameters the NS component then makes up $56^{+23}_{-35}$\% of that flux.  Torres et al's method is common in the literature \citep[e.g.][]{Campana02,Campana05b}, but we do not feel it is the most appropriate when constraining the contributions of broad band components.  

\begin{figure}
\figurenum{4}
\epsscale{0.6}
\includegraphics[angle=0,scale=.5]{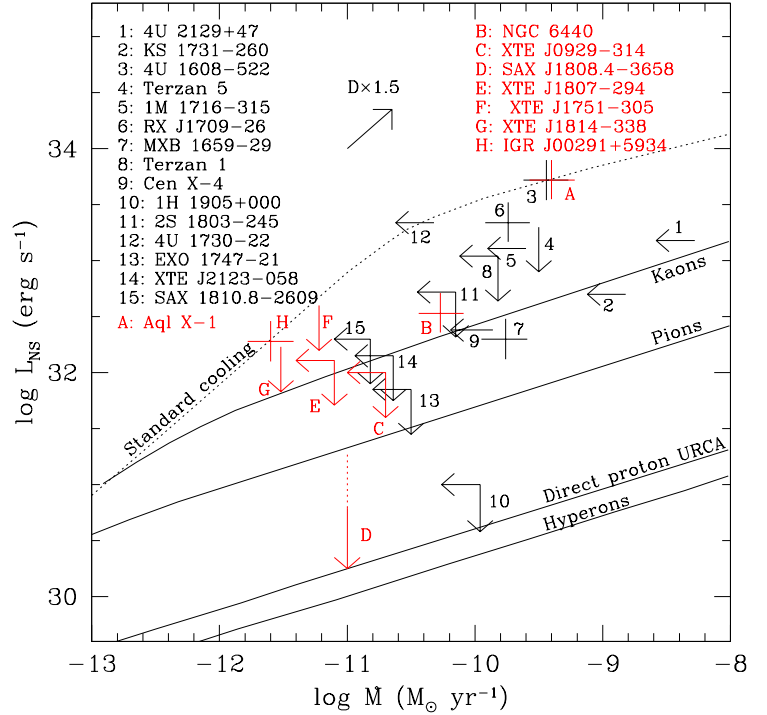}
\caption[cooling.eps]{ 
Measurements of, or limits on, the quiescent thermal luminosity of various NS transients, 
compared to estimates of, or upper limits on, their time-averaged mass transfer rates.  
Data from the compilations of \citet{Heinke07a} and Table 3.  
The predictions of standard NS cooling and 
various enhanced cooling mechanisms are plotted, following \citet{Yakovlev04}. 
The accreting millisecond pulsars (including the intermittent ones) are labeled separately (in red).  The effect of using a MEKAL rather than a power-law continuum for SAX J1808.4-3658, allowing a higher upper limit on the NS component, is indicated with the dotted line.  The effect of increasing the distance by a factor of 1.5 for any system is indicated with an arrow (labeled Dx1.5).  
\label{fig:cooling}} 
\end{figure}

The best estimates of the mass transfer rates and quiescent X-ray luminosities in Table \ref{tab:mdots} 
are plotted in Figure \ref{fig:cooling}.  These estimates suffer uncertainties in distances and recurrence times (where appropriate, we plot upper limits, in some cases upper limits in both mass transfer rate and $L_{NS}$).  However, there 
is clearly a large variation in cooling behavior exhibited by these accreting neutron stars, with quiescent 
thermal luminosities ranging by several orders of magnitude. \citet{Heinke07a} suggested that NSs with lower mass transfer rates, which are generally older systems, have experienced significant mass transfer over their lives and are now more massive, and inclined to faster cooling.  The larger dataset collected here argues against this idea.  In addition to the two systems with high mass transfer rates, 4U 1730-22 and 00291 lie near the predictions of standard cooling.  We note the contrast between the bright thermal component of 00291  and the faint thermal component of 1808.  These two systems have similar periods (2.46 and 2.01 hours, respectively) and heated low-mass (brown dwarf) companions \citep{Bildsten01,Galloway05}.  One might expect similar evolutionary histories, similar amounts of accreted mass, and thus similar cooling rates from these two systems.  
The observed difference in cooling requires different neutrino emission mechanisms, suggesting that 1808's NS may be significantly more massive than that in IGR J00291+5934.   
This might be interpreted as due to differences in their mass transfer histories, or by the NS in 1808 being born with a higher mass.  

There are a few caveats to consider.  The evidence for the thermal component in 00291 is only at the 3$\sigma$ level.  It is not absolutely certain that the thermal component in 00291 is produced from deep crustal heating, especially considering the odd quiescent X-ray behavior seen by \citet{Jonker05}.  The distance and mass transfer rate of 00291 are not known to high accuracy.  However, changes in 00291's distance by a factor of two, or mass transfer rate by a factor of 10 or more, would not eliminate the contrast between 1808 and 00291 (see Fig. 4).

\section{Conclusions}

The three combined XMM observations of SAX J1808.4-3658 in quiescence provide one of the most stringent constraints on the thermal component of any transiently accreting neutron stars observed so far, with the thermal $L_{NS,bol}<6\times10^{30}$ ergs/s (for a powerlaw continuum model) or $<1.9\times10^{31}$ ergs/s (for a MEKAL thermal plasma model).  Combined with 1808's well-constrained mass transfer rate and distance, this constraint strongly requires enhanced neutrino cooling from the NS in 1808.  The models of kaon and pion cooling presented by \citet{Yakovlev04} are excluded by this constraint, favoring direct  URCA neutrino emission processes involving protons, hyperons or deconfined quarks, depending on the constituents of matter at supra-nuclear densities.  

We have compiled literature estimates of the mass transfer rates and quiescent thermal luminosities of a large number of transient NS LMXBs.  Uncertainties in transient outburst histories and the lack of detections of thermal components limit the number of useful data points.  Although many of the measurements are upper limits, there is strong evidence for a range of NS neutrino cooling rates.  Uncertainties in the distances, while significant for many systems, affect quiescent and outburst flux equally (see Fig.~4), and thus do not greatly affect these results.  

The accreting millisecond pulsar IGR J00291+5934 shows evidence for a thermal component in its quiescent X-ray spectrum, which is consistent with the prediction from standard cooling.  00291 seems similar to 1808 (both show heated brown dwarf companions, and expanding $\sim$2-hour orbits).  Thus 00291's much hotter NS (implying a less massive NS) comes as a surprise.  If 00291's thermal emission is a result of deep crustal heating, then it is likely that 00291 either experienced a different binary evolution than 1808 (e.g. starting with a smaller-mass companion?), or began its evolution with a smaller-mass NS.

{\it Note added in proof}--\citet[submitted]{Campana08b} have independently found evidence for a thermal component in 00291 from XMM data, confirming our detection of such a component.

\acknowledgements

We thank M. Prakash, D. Page, K. Levenfish, D. Blaschke, and S. Campana for discussions, and the anonymous referee for a useful report.  
 COH acknowledges support from the Lindheimer Postdoctoral Fellowship at 
Northwestern University, NASA \Chandra\ grants G07-8078X, G08-9053X, and G08-9085X, and NASA XMM grants NNX06AE78G and NNX06AH62G. CJD acknowledges support from NASA XMM grant NXX06AH62G and \textit{Chandra} grant TM7-8007X.  PGJ acknowledges support from the Netherlands Organization for Scientific Research.  Support for this work is provided in part by the Theoretical Institute for Advanced Research in Astrophysics
   (TIARA) operated under Academia Sinica and the National Science
   Council Excellence Projects program in Taiwan administered through
   grant number NSC 96-2752-M-007-007-PAE.



\bibliography{src_ref_list}
\bibliographystyle{apj}


\begin{deluxetable}{ccccccc}
\tablewidth{6.0truein}
\tablecaption{\textbf{Spectral Fits to SAX J1808.4-3658}}
\tablehead{
\colhead{\textbf{Epoch}}  & $N_H\times10^{21}$ & $\Gamma$/kT  & $L_X$ & $\chi^2_{\nu}$/dof  & NS kT, eV & $L_{NS,bol}$  \\
}
\startdata
\hline \\
\multicolumn{7}{c}{Power-law + NSATMOS Fit} \\
\hline \\
2001 & (1.3) & $1.61^{+0.20}_{-0.21}$ & 7.6$^{+1.2}_{-1.4}\times10^{31}$ & 0.83/83 & $<30$ & $<6.2\times10^{30}$  \\ 
2006 & ... & $1.83^{+0.16}_{-0.17}$ & 5.1$^{+0.5}_{-0.7}\times10^{31}$ & ... & ... & ... \\ 
2007 & ... & $1.74^{+0.11}_{-0.12}$ & 7.9$\pm0.7\times10^{31}$ & ... & ... & ... \\ 
\hline \\
\multicolumn{7}{c}{MEKAL + NSATMOS Fit} \\
\hline \\
2001 & (1.3) & $8^{+21}_{-3}$ & 7.5$^{+2.4}_{-1.2}\times10^{31}$ & 0.90/83 & $36^{+4}_{-8}$ & $1.3^{+0.6}_{-0.8}\times10^{31}$  \\ 
2006 & ... & $5.4^{+2.9}_{-1.4}$ & 5.1$^{+0.5}_{-0.7}\times10^{31}$ & ... & ... & ... \\ 
2007 & ... & $5.4^{+1.9}_{-1.2}$ & 7.6$\pm0.7\times10^{31}$ & ... & ... & ... \\ 
\enddata
\tablecomments{Simultaneous spectral fits to SAX J1808.4-3658 data, from 2001, 2006 and 2007, to either a power-law + NSATMOS model, or MEKAL + NSATMOS model.  
Parameters and results from the fit as a whole are listed only in the first line for each fit.   
Errors are 90\% confidence for a single parameter.  
$N_H$ is held fixed at the interstellar value, in agreement with Swift measurements in outburst (see text). 
Luminosities in erg s$^{-1}$; $L_X$ for 0.5-10 keV, $L_{NS,bol}$ for 0.01-10 keV. 
\label{tab:1808spec}}
\end{deluxetable}

\begin{deluxetable}{cccccccccc}
\tabletypesize{\footnotesize}
\tablewidth{7.2truein}
\tablecaption{\textbf{Estimates of Quiescent NS Luminosities and Mass Transfer Rates}}
\tablehead{
\colhead{\textbf{Source}}  & $P_{orb}$ & $N_H$ & $kT$ & $D$ & Outbursts & Timeline  & \.{M} & $L_{NS}$ & Refs \\
  & (Hours) &  ($10^{21}$ cm$^{-2}$) & (eV) & (kpc) & & (years) & (\Msun yr$^{-1}$) & (erg s$^{-1}$) & \\
}
\startdata
SAX J1808-3658 & 2.01 & 1.3 & $<30$ & 3.5 & 5 & 12 & $9\times10^{-12}$ & $<4.9\times10^{30}$ & 1,2,3 \\
1H 1905+000      & $<1.5$? & 2.1 & $<$40 & 10  & 1 & 34 & $<1.1\times10^{-10}$ & $<1.0\times10^{31}$ & 4,5 \\
2S 1803-45       & $\sim9$? & 1.47 & $<$91 & 7.3 & 2 & 33 & $<7\times10^{-11}$ & $<5.2\times10^{32}$  & 6,2 \\
1M1716-315       & $<1.5$? & 7.4 & 116 & 6.9 & 1 & 37 & $<2.5\times10^{-10}$ & $1.3\times10^{33}$ & 7,1 \\
4U 1730-22       & - & 3.7  & 131  & 10 & 1 & 37 & $<4.8\times10^{-11}$ & $2.2\times10^{33}$ & 8,9,1 \\
4U 2129+47       & 5.24 & 1.7  & 119 & 6.3 & 1 & 45 & $<5.2\times10^{-9}$ & $1.5\times10^{33}$ & 10,11,12,1 \\
XTE J1751-305    & 0.71 & 9.8 & $<71$ & 8 & 4 & 12 & $6\times10^{-12}$ & $<4\times10^{32}$  & 13,14,15,1,2 \\
XTE J0929-314    & 0.73 & 0.76 & $<50$ & 10/9 & 1 & 12 & $<2.0\times10^{-11}$ & $<1.0\times10^{32}$ & 16,17,18,15,19 \\
XTE J1807-294    & 0.67 & 4.6 & $<51$ & 8  & 1 & 12 & $<8\times10^{-12}$ & $<1.3\times10^{32}$ &  20,17,19,2 \\
XTE J1814-338    & 4.27 & 1.6 & $<69$ & 8 & 2 & 19 & $3\times10^{-12}$ & $<1.7\times10^{32}$  & 21,22,17,2 \\
IGR 00291+5934  & 2.46 & 2.8 & 71 & 4 & 3 & 12 & $2.5\times10^{-12}$ & $1.9\times10^{32}$  & 23,24,25,2 \\
\enddata
\tablecomments{Estimates of quiescent thermal NS luminosities from X-ray transients, orbital periods (those followed by ? are rough estimates--see the references), assumed distances, and mass transfer rates (inferred from RXTE ASM observations for systems with RXTE-era outbursts, best historical data otherwise).  Quiescent thermal luminosities are computed for the unabsorbed NS component in the 0.01-10 keV range.  Outbursts and years columns give the number of outbursts seen and the time baseline used to compute \.{M}.  References as follows:
1: Mass transfer rate computed in this work,
2: Quiescent NS thermal luminosity computed in this work,
3: \citet{Galloway06},
4: \citet{Jonker06},
5: \citet{Jonker07a},
6: \citet{Cornelisse07},
7: \citet{Jonker07b},
8: \citet{Tomsick07},
9: \citet{Chen97},
10: \citet{Nowak02},
11: \citet{Pietsch86},
12: \citet{Wenzel83},
13: \citet{Markwardt02},
14: \citet{Miller03},
15: \citet{Wijnands05b},
16:  \citet{Galloway02},
17: \citet{Galloway06b},
18: \citet{Juett03},
19: \citet{Campana05b},
20: \citet{Campana03}, 
21: \citet{Krauss05},
22: \citet{WijRey03},
23: \citet{Galloway05},
24: \citet{Jonker05},
25: \citet{Torres08}
\label{tab:mdots}
}
\end{deluxetable}

\begin{deluxetable}{ccccccc}
\tablewidth{5.6truein}
\tablecaption{\textbf{Spectral Fits to IGR J00291+5934}}
\tablehead{
\colhead{\textbf{Model}}  & $N_H\times10^{21}$ & $\Gamma$  
& $L_X$ & $\chi^2_{\nu}$/dof  & kT, eV 
& $L_{NS,bol}$  \\
}
\startdata
PL    & (4.6) & 2.6$^{+0.5}_{-0.4}$ & $1.3\times10^{32}$ & 1.08/12 & - & - \\
NS+PL & (4.6) & 1.5$\pm0.7$ & $1.8\times10^{32}$ & 0.54/11 & 71$^{+5}_{-9}$ & 1.9$^{+0.6}_{-0.8}\times10^{32}$ \\
\enddata
\tablecomments{Spectral fits to 2005 \Chandra\ data on IGR J00291+5934.  A distance of 4 kpc is assumed, and $N_H$ is fixed to the value determined by \citet{Torres08}.  Luminosities in erg s$^{-1}$; $L_X$ for 0.5-10 keV, $L_{NS,bol}$ for 0.01-10 keV. 
\label{tab:00291spec}}
\end{deluxetable}


\end{document}